\documentclass[pra,twocolumn,showpacs]{revtex4}
\usepackage[dvips]{graphicx}
\usepackage{amsmath,amsfonts,amssymb}
\usepackage{dcolumn}
\usepackage{bm}
\usepackage{color}
\usepackage{epsfig}
\usepackage{hyperref}
\begin{document}

\def\makered#1{{\color{red} #1}}
\def\jpb{J. Phys. B: At. Mol. Opt. Phys.~}
\def\pra{Phys. Rev. A~}
\def\prb{Phys. Rev. B~}
\def\prl{Phys. Rev. Lett.~}
\def\jmo{J. Mod. Opt.~}
\def\jetp{Sov. Phys. JETP~}
\def\etal{{\em et al.}}

\def\reff#1{(\ref{#1})}

\def\tsim{t_\mathrm{sim}}
\def\diff{\mathrm{d}}
\def\Re{\mathrm{Re}\,}
\def\Ncyc{N_\mathrm{cyc}}
\def\imagi{\mathrm{i}}

\def\halb{\frac{1}{2}}

\def\pabl#1#2{\frac{\partial #1}{\partial #2}}

\def\beq{\begin{equation}}
\def\eeq{\end{equation}}

\def\beqa{\begin{eqnarray}}
\def\eeqa{\end{eqnarray}}
\def\energy{{\cal E}}
\def\energykin{{\cal E}_\mathrm{kin}}

\def\eulere{\mathrm{e}}

\def\alphahat{\hat{\alpha}}
\def\Ehat{\hat{E}}
\def\Ahat{\hat{A}}

\def\ket#1{\vert #1\rangle}
\def\bra#1{\langle#1\vert}
\def\braket#1#2{\langle #1 \vert #2 \rangle}

\def\vekt#1{\bm{#1}}
\def\vect#1{\vekt{#1}}
\def\vektr{\vekt{r}}

\def\makered#1{{\color{red} #1}}

\def\Im{\,\mathrm{Im}\,}

\def\varphic{\varphi_{\mathrm{c}}}

\def\Up{U_\mathrm{p}}
\def\vxc{v_\mathrm{xc}}
\def\vextop{\hat{v}_\mathrm{ext}}
\def\VC{V_\mathrm{c}}
\def\VHX{V_\mathrm{Hx}}
\def\VHXC{V_\mathrm{Hxc}}
\def\wop{\hat{w}}
\def\Gammaevenodd{\Gamma^\mathrm{even,odd}}
\def\Gammaeven{\Gamma^\mathrm{even}}
\def\Gammaodd{\Gamma^\mathrm{odd}}

\def\hamop{\hat{{\cal H}}}
\def\hamopKH{\hat{{\cal H}}_\mathrm{KH}}
\def\Hop{\hat{H}}
\def\Gop{\hat{G}}
\def\GKH{\hat{G}_\mathrm{KH}}
\def\Pop{\hat{P}}
\def\pop{\hat{p}}
\def\HopKS{\hat{H}_\mathrm{KS}}
\def\HKS{H_\mathrm{KS}}
\def\Top{\hat{T}}
\def\TopKS{\hat{T}_\mathrm{KS}}
\def\VopKS{\hat{V}_\mathrm{KS}}
\def\VKS{{V}_\mathrm{KS}}
\def\vKS{{v}_\mathrm{KS}}
\def\Ttildeop{\hat{\tilde{T}}}
\def\Ttilde{{\tilde{T}}}
\def\Vextop{\hat{V}_{\mathrm{ext}}}
\def\Vext{V_{\mathrm{ext}}}
\def\Vopee{\hat{V}_{{ee}}}
\def\psiopdag{\hat{\psi}^{\dagger}}
\def\psiop{\hat{\psi}}
\def\vext{v_{\mathrm{ext}}}
\def\Vee{V_{ee}}
\def\vee{v_{ee}}

\def\Psigrid{\Psi_{\#}}

\def\nop{\hat{n}}
\def\Uop{\hat{U}}
\def\Wop{\hat{W}}
\def\bop{\hat{b}}
\def\bopdag{\hat{b}^{\dagger}}
\def\qop{\hat{q}}
\def\jop{\hat{j\,}}
\def\vHxc{v_{\mathrm{Hxc}}}
\def\vHx{v_{\mathrm{Hx}}}
\def\vH{v_{\mathrm{H}}}
\def\vc{v_{\mathrm{c}}}
\def\xop{\hat{x}}
\def\\{\par}
\def\varphiexact{\varphi_{\mathrm{exact}}}

\def\fmathbox#1{\fbox{$\displaystyle #1$}}

\title{Floquet analysis of real-time wavefunctions without solving the Floquet equation}

\author{V.\ Kapoor}
\affiliation{Institut f\"ur Physik, Universit\"at Rostock, 18051 Rostock, Germany}

\author{D.\ Bauer}
\affiliation{Institut f\"ur Physik, Universit\"at Rostock, 18051 Rostock, Germany}

\date{\today}

\begin{abstract}
 We propose a method to obtain Floquet states---also known as light-induced states---and their quasi-energies from real-time wavefunctions without solving the Floquet equation. This is useful for the analysis of various phenomena in time-dependent quantum dynamics if the Hamiltonian is not strictly periodic, as in short laser pulses, for instance. There, the population of the Floquet states depends on the pulse form and is automatically contained in the real-time wavefunction but not in the standard Floquet approach. Several examples in the area of intense laser-atom interaction are exemplarily discussed: (i) the observation of even harmonics for an inversion-symmetric potential with a single bound state; (ii) the dependence of the population of Floquet states on (gauge) transformations and the emergence of an invariant, observable photoelectron spectrum; (iii) the driving of resonant transitions between dressed states, i.e.,  the dressing of dressed states, and (iv) spectral enhancements at channel closings due to the ponderomotive shift of above-threshold ionization peaks.
\end{abstract}
\maketitle

\section{Introduction}
The time-dependent Schr\"odinger equation (TDSE) with a time-periodic Hamiltonian has solutions which can be expressed in a time-periodic basis. This basis is referred to as the Floquet basis, and eigenstates in this basis are the Floquet states \cite{floquet,floquetclassics,floquetinbooks,floquetreview}. Time-periodic potentials naturally arise when matter is exposed to laser fields.  In this context, Floquet states are also known as ``light-induced states'' (LIS) \cite{potvliegeLIS}, as they are the new states of the combined system ``target + laser field''. In fact, Floquet theory has been used to determine, e.g., very accurate ionization rates \cite{potvliegeshakeshaft,potvliegecode}. Using so-called $R$-matrix Floquet theory, the method has been extended to multi-electron systems \cite{rmatrixfloquet}. Strict periodicity of the Hamiltonian with the laser period implies physically  that the laser pulse was always on and will be on forever. Then the problem arises how the field-free system under study, e.g., an atom, gets into the laser field in the first place and how the field-free observables emerge. In fact, the population of the Floquet states depends on the laser pulse form. If the (up and down) ramping of the laser field is adiabatic and the laser frequency is non-resonant we expect the system to follow just a single Floquet state, namely the one which is adiabatically connected to the field-free initial state. However, for non-adiabatic ramping or resonant interactions a superposition of Floquet states is created. An example for non-adiabatic population of several Floquet states, {leading to an apparent generation of even harmonics in} inversion-symmetric potentials, is given in Sec.~\ref{harmonics} of this paper.

Instead of converting the TDSE into the time-independent Floquet equation [eq.~\reff{flqeq} below]  one  may alternatively solve it directly in real-time. In the latter case there are no assumptions about periodicity or adiabatic ramping and, e.g.,  the effect of different laser pulse forms can be studied. However, the direct solution of the TDSE in real-time does not involve the Floquet basis so that information about LIS are not directly available. As many interesting phenomena such as the AC Stark effect, Rabi oscillations, or stabilization against ionization \cite{gavrila,grobefedorov} is most conveniently analyzed in terms of LIS, it is desirable to extract the ``Floquet information'' from the real-time wavefunction ``on-the-fly'' while propagating (or by post-processing) it, without having to solve the Floquet equation as well.       
We present such a method to analyze non-perturbative, laser-driven quantum dynamics via the (time-resolved) Floquet information contained in the corresponding real-time wavefunction.

The paper is organized as follows: in Sec.~\ref{preliminaries} we review the basics of Floquet theory.  In Sec.~\ref{harmonics} we briefly summarize the general derivation of harmonic generation selection rules before we present the (at first sight surprising) {presence of peaks at even harmonics of the laser frequency} in the case of an inversion-symmetric potential with only one bound state.  In Sec.~\ref{method} we introduce our method to obtain the Floquet information from the real-time wavefunction, e.g., the populated states and their energies, and use them to explain the presence of {hyper-Raman lines at even harmonic frequencies.}  In Sec.~\ref{trafos} we investigate how the population of Floquet states changes under (gauge) transformations while the Floquet energies and the observable photoelectron spectra remain invariant.  In Sec.~\ref{photoelspec}  time-resolved Floquet spectra of real-time wavefunctions in the so-called velocity gauge and in the Kramers-Henneberger frame-of-reference are compared. In Sec.~\ref{channelclose} the channel-closing phenomenon and related spectral enhancements are interpreted in terms of Floquet state-crossings. A conclusion is given in  Sec.~\ref{sumout}. In this work we restrict ourselves to spatially one-dimensional (1D) model Hamiltonians. It is straightforward to extend the method to higher dimensions, as indicated in Appendix~\ref{3D}.  Atomic units (a.u.) $\vert e\vert = m_e = \hbar = 4\pi\epsilon_0=1$ are used unless noted otherwise.

\section{Basic theory} \label{preliminaries}
Consider a linearly polarized laser field $E(t)$ of frequency $\omega_1$ in dipole approximation, polarized along the $x$-direction and interacting with an electron in some binding potential $V$. The Hamiltonian in length gauge reads
\beq \Hop(t)=\Hop_0+\Wop(x,t), \qquad \Wop(x,t)=x E(t) \label{hamiltonian0} \eeq
with 
\beq \Hop_0=-\frac{1}{2}\pabl{^2}{x^2}+V(x). \label{Hnull}\eeq

\subsection{Floquet theory}
For sufficiently long laser pulses 
\beq E(t+T)= E(t),\qquad T=\frac{2\pi}{\omega_1} \label{intperiod} \eeq
holds to high accuracy,  and thus also $\Wop(t+T) = \Wop(t)$ so that
\beq \Hop(t+T) = \Hop(t) . \eeq
The Floquet theorem \cite{floquet,floquetclassics,floquetinbooks,floquetreview} states  that in this case the TDSE
\beq\imagi\pabl{}{t}\Psi(x,t) = \Hop(t) \Psi(x,t) \label{TDSE} \eeq
 has solutions of the form
\beq \Psi(x,t)=\eulere^{-\imagi\epsilon t} \Phi(x,t), \label{psixt}\eeq
 $\Phi(x,t)$ being periodic itself, 
\beq \Phi(x,t)=\Phi(x,t+T). \label{periodicphi}\eeq
$\epsilon$ is called the quasienergy or Floquet energy.
The wavefunctions $\Phi(x,t)$ fulfill the Schr\"odinger equation
\beq \hamop(t) \Phi(x,t) = \epsilon\Phi(x,t) \label{fhami}\eeq
with 
\beq \hamop(t) = \Hop(t) - \imagi\pabl{}{t}. \label{ffham}\eeq
If $\epsilon$ is an eigenvalue and $\Phi(x,t)$ the corresponding eigenstate, also 
\beq \epsilon'=\epsilon+m\omega_1,\quad \Phi'(x,t)=\eulere^{\imagi m \omega_1 t} \Phi(x,t),\quad m\in\Bbb{Z} \eeq
are solutions of \reff{fhami}. Owing to the time periodicity of $\Phi(x,t)$ we can expand
\beq \Phi(x,t)=\sum_{n=-\infty}^\infty \varphi_n(x) \eulere^{-\imagi n\omega_1 t}\label{fourexp}.\eeq 
For a monochromatic laser field the interaction Hamiltonian $\Wop(x,t)$ can be written as
\beq \Wop(x,t)=\Wop^{+}(x)\exp(\imagi \omega_1 t) +\Wop^{-}(x)\exp(-\imagi \omega_1 t), \eeq 
leading to the time-independent Floquet equation 
\begin{eqnarray} \lefteqn{(\epsilon +n\hbar\omega_1 -\Hop_0)\varphi_n(x)} \label{flqeq} \\ & = & \Wop^{+}(x)\varphi_{n+1}(x)+ \Wop^{-}(x)\varphi_{n-1}(x). \nonumber  \end{eqnarray}
The index $n$ of the Floquet state is known as the ``block index,'' which may be interpreted as the number of photons involved in the process under study. Hence, the Floquet equation \reff{flqeq}  couples any Floquet block $n$ with its neighboring blocks $n\pm 1$ via absorption or emission of a photon. 

{In principle,  \reff{flqeq}  is an infinite-dimensional set of differential equations.} In practice,  it is truncated  so that $n_{\min} \leq n \leq n_{\max}$. 
In obtaining the eigenvalue equation \reff{flqeq}, we assumed strict time-periodicity, which physically means that the laser pulse is always on.

\subsection{Non-Hermitian Floquet Theory}    
{ We are interested in systems which, in the field-free situation, possess besides bound states also a continuum. In the presence of a laser field such a system may ionize, i.e., the field-free stationary states are turned into field-dressed, quasi-stationary states. The simplest cases of only a few (field-free) bound states (allowing for resonances) plus a continuum dressed by laser fields have been discussed in the literature since long ago (see, e.g., \cite{kazakov} and \cite{fedorovreview} for a review). In an actual implementation of Floquet theory, the decay of  quasi-stationary states needs to be taken into account when solving  \reff{flqeq} by applying Siegert boundary conditions for the outgoing waves \cite{potvliegecode}, leading to complex Floquet energies
\beq \epsilon=\Re \epsilon - \imagi \frac{\Gamma}{2} \label{complexfloquenergs} \eeq
where $\Gamma$ is the ionization rate. The difference between $\Re \epsilon$ and the field-free $\epsilon^{(0)}$ is the AC Stark shift.}

\subsection{Finite-grid, finite-pulse TDSE solution} 
{ We solve
\beq\imagi\pabl{}{t}\Psigrid(x,t) = \Hop(t) \Psigrid(x,t) \label{TDSEgrid} \eeq
 on a numerical grid of size $L$,  $-\frac{L}{2}< x < \frac{L}{2}$ for times $0<t<\tsim$ with $\tsim$ the total simulation time.  The binding potential $V(x)$ is centered at $x=0$. In all cases discussed in this work we start from the field-free ground state on the grid $\Psigrid(x,0)=\Psigrid^{(0)}(x)$. Probability density approaching the grid boundary is absorbed by an imaginary potential. } 

{
Our aim in the following will be to analyze $\Psigrid(x,t)$ in terms of Floquet energies and states.}

\section{Harmonic generation} \label{harmonics}
{In the first example we apply our method to investigate the origin of {apparently even harmonics}  in an inversion-symmetric potential with only one bound state.}

There are many ways to derive selection rules for harmonic generation (HG). Most elegant, rigorous, and appropriate for our purpose is the approach employing dynamical symmetries  \cite{HGselec,cecch}. 
Consider the stationary Schr\"odinger equation
\beq \Hop_0\Psi(x) = \energy\Psi(x), \label{statSE}\eeq
with $\Hop_0$ given by \reff{Hnull}.
If the potential $V$ is inversion-symmetric, $V(x)=V(-x)$, the Hamiltonian $\Hop_0$ is invariant under spatial inversion as well, 
\beq \Pop_p f(x) = f(-x),\quad \Pop_p^2=1, \ \Pop_p^{-1}=\Pop_p, \eeq
\beq [\Hop_0,\Pop_p]=0, \eeq
so that for non-degenerate energies $\energy$ the eigenstate  $\Psi(x)$ is also an eigenstate of the spatial-inversion operator $\Pop_p$. Because of $\Pop_p^2=1$ the eigenvalues can only be $\pm 1$ (parity):
\beq \Pop_p \Psi(x) = \pm \Psi(x). \eeq

The full Hamiltonian \reff{hamiltonian0} and  the Floquet-Hamiltonian  \reff{ffham}  are not invariant under spatial inversion but under the dynamical symmetry operation ``spatial inversion combined with a translation in time by half a period'',
\beq [\Hop(t),\Pop_{pt}]=[\hamop(t),\Pop_{pt}]=0, \eeq
\beq \Pop_{pt} f(x,t)= f(-x,t+\pi/\omega_1),\qquad \Pop_{pt}^2=1, \label{poppt} \eeq
\beq \Pop_{pt}=\Pop_p \Pop_t=\Pop_t \Pop_p, \qquad \Pop_t f(x,t)=  f(x,t+\pi/\omega_1). \eeq
For non-degenerate $\epsilon$ 
\beq \Pop_{pt} \Phi(x,t) = \pm \Phi(x,t).  \label{symmPpt}\eeq  
Because of \reff{fourexp} we observe that
\beq \Pop_{pt}\Phi(x,t)=\sum_{n} (-1)^n \eulere^{-\imagi n\omega_1 t} \Pop_p \varphi_n(x), \label{alternparity}\eeq
and with \reff{symmPpt} follows that 
\beq  \Pop_p \varphi_n(x)= \pm (-1)^n  \varphi_n(x),\label{parity}\eeq 
i.e., the $\varphi_n(x)$ have an alternating parity with respect to the Floquet block index $n$. 

{Numerically, the HG spectrum $\sim \omega^4 \vert d(\omega)\vert^2$ is calculated via the Fourier-transformed dipole moment
\beq d(\omega)\sim \int_{0}^{\tsim} \diff t    \int_{-\frac{L}{2}}^{\frac{L}{2}}\!\!\diff x\,  \Psigrid^*(x,t) x \Psigrid(x,t)\, \eulere^{\imagi\omega t}. \eeq
Assuming that the numerically determined, exact wavefunction on the grid is well described by just a single Floquet state, 
using \reff{psixt}, \reff{fourexp}, and \reff{complexfloquenergs} yields
\beq d(\omega)\sim     \sum_{nm} \int_{-\frac{L}{2}}^{\frac{L}{2}} \varphi_m^*(x) x \varphi_n(x)\, \diff x \qquad \qquad \label{dw1} \eeq
\[  \qquad \qquad\qquad \qquad \times \int_0^{\tsim} \eulere^{t\{\imagi [\omega-\omega_1(n-m)]-\Gamma \} } \,\diff t. \]}
The spatial integral is non-vanishing only if $\varphi_n$ has the opposite parity of $\varphi_m$, i.e.,
\beq n-m = 2k+1,\qquad k \in \Bbb{Z} .\eeq
{The temporal integral thus leads to peaks centered at frequencies
\beq \omega=(2k+1)\omega_1 \label{selecrule1} \eeq
with widths determined by $\tsim$ (frequency-time uncertainty) and $\Gamma$ (decay).}
The selection rule \reff{selecrule1} is the well-known result that an inversion-symmetric target   in a linearly polarized laser field  generates {\em odd} harmonics only. Note that the above derivation also holds for multi-electron targets because the electron-electron interaction is also invariant under the symmetry operations $\Pop_{p}$ and  $\Pop_{pt}$.

\begin{figure}
\includegraphics[width=0.45\textwidth]{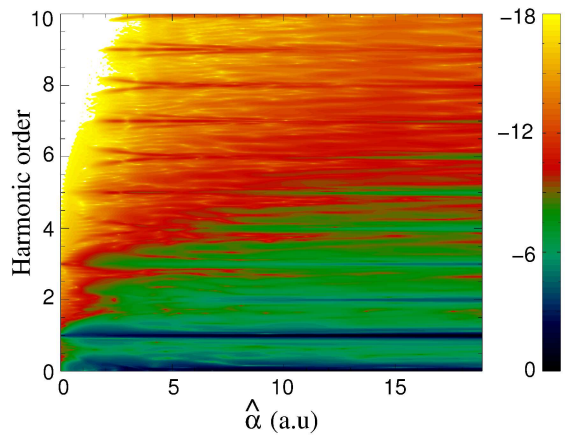} 
\caption{ (color online). Logarithmically scaled HG strength $\omega^4 \vert d(\omega)\vert^2$  vs harmonic order and excursion amplitude $\alphahat=\Ahat/\omega_1$ ($\omega_1=1$, vector potential $A(t)$ ramped up and down over 4 cycles and held constant with amplitude $\Ahat$ for 30 cycles). {The numerical fast-Fourier transform was performed over the pulse duration, i.e., $\tsim=38$\,cycles, using a Hanning window.}  \label{fig:harm1}}
\end{figure}

\subsection{Hyper-Raman lines at even harmonics of the laser frequency}
It is known that HG peaks at positions different from odd multiples of the fundamental laser frequency $\omega_1$ are to be expected for an inversion-symmetric potential if at least two Floquet states of opposite parity are populated \cite{bavlimetiu,moiseyevlein}. Physically, the superposition of two Floquet states may amount to, e.g., the absorption of $n$ photons of energy $\omega_1$ but emission of one photon of energy $n\omega_1-\Delta \epsilon$, with $\Delta \epsilon$ being the energy difference between initial and final state. This should lead to hyper-Raman lines in the spectra which, however, are typically weak \cite{dipiazza,moiseyevlein}. Nevertheless, {if observable, they appear at even harmonics of the laser frequency in the case of degeneracy, $\Delta \epsilon=0$.}  

We consider an electron in  the P\"oschl-Teller potential
\beq V(x) = - \frac{1}{\cosh^2 x} \label{ptpot}\eeq
and subject to a laser field. The  potential \reff{ptpot} supports only a single bound state $\Psi_0(x)$  of energy $\energy_0=-0.5$. Hence, superpositions of field-free bound states are ruled-out. As a consequence, perturbation theory in the external field can certainly not predict hyper-Raman lines or even harmonics. However, Fig.~\ref{fig:harm1} shows the logarithmically scaled HG strength $\omega^4 \vert d(\omega)\vert^2$  as obtained from the numerical solution of the TDSE. The HG strength is plotted vs harmonic order $\omega/\omega_1$  and the amplitude $\alphahat$ of the excursion 
\beq \alpha(t)=\int^t A(t)\,\diff t  \label{alphaintegral} \eeq
with $A(t)$ the vector potential of the laser field. The electric field is given by $E(t)=-\partial_t A(t)$. Given the vector potential amplitude $\Ahat$, the excursion amplitude is $\alphahat=\Ahat/\omega_1$, the field amplitude $\Ehat=\Ahat\omega_1$. The laser pulse parameters are specified in the figure caption. One sees that for sufficiently strong excursion amplitude $\alphahat$ { peaks at even harmonics of the laser frequency} appear too. Picking an even harmonic at $\alphahat>15$ (e.g., the 6th) and tracing it back to low $\alphahat$ reveals that the peak  splits and rapidly drops in magnitude (e.g., around $\alphahat\simeq 2$ for the 6th harmonic). In the next Section we will use our real-time Floquet method to show that the appearance of even harmonics is due to the population of several LIS that become quasi-degenerate as $\alphahat$ increases.

\subsection{Superposition of Floquet states}
In  the case of a non-adiabatic transfer of the field-free state to field-dressed states one has to allow for a superposition of Floquet states {in order to represent the exact, numerically determined wave function on the grid,}
\beq \Psigrid(x,t)\simeq\sum_\beta \eulere^{-\imagi\epsilon_\beta t} \Phi_\beta(x,t)=\sum_{\beta n} \eulere^{-\imagi t(\epsilon_\beta +n\omega_1)} \varphi_{\beta n}(x). \label{psixtsuperpos}\eeq
{Here we assume that the expansion coefficients are included in $\Phi_\beta(x,t)$ and $\varphi_{\beta n}(x)$. For continuous quasienergies the sum over $\beta$ should be replaced by an integral over $\epsilon$. The Fourier-transformed dipole will be
\beq d(\omega) \sim      \sum_{\beta\gamma nm} \int_{-\frac{L}{2}}^{\frac{L}{2}} \varphi_{\gamma m}^*(x) x \varphi_{\beta n}(x)\, \diff x \qquad \qquad\ \label{dw2}  
\eeq
\[  \qquad \times \int_0^{\tsim} \eulere^{t\{\imagi [\omega-\omega_1(n-m) - (\Re\epsilon_\beta-\Re\epsilon_\gamma)]-(\Gamma_\beta+\Gamma_\gamma)/2 \} } \,\diff t. \]}

Again, in order for the spatial integral to not vanish the parity of $\varphi_{\beta n}$ and $\varphi_{\gamma m}$ must be different. However, now this can be the case not only for $n-m=2k+1$, but also for  $n-m=2k$ if the parity of, e.g., $\varphi_{\beta 0}$ is opposite to the one  of $ \varphi_{\gamma 0}$.
Hence, one expects the above-mentioned hyper-Raman peaks at
\beq \omega=k\omega_1 +\Delta\epsilon,\qquad k\in\Bbb{Z} \label{peakpos}\eeq
where {$\Delta\epsilon=\Re\epsilon_\beta-\Re\epsilon_\gamma$ is the difference between the real parts of the two Floquet quasienergies involved.} Thus, in order to observe even harmonics at exactly $\omega=2k \omega_1$ a degeneracy {$\Re\epsilon_\beta=\Re\epsilon_\gamma$} is required. Such a degeneracy between the (field-dressed) initial state and another one of opposite parity is also likely to populate the latter one.

\begin{figure}
\includegraphics[width=0.45\textwidth]{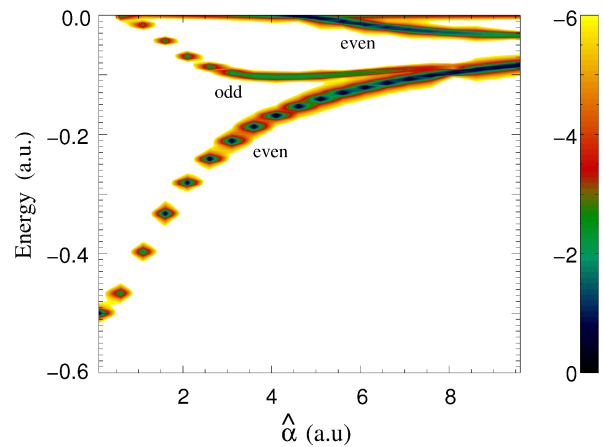} 
\caption{ (color online). Logarithmic plot of $ R = \vert Q_+\vert^2 + \vert Q_- \vert^2 $ vs energy $\energy$ and excursion amplitude $\hat{\alpha}=\Ahat/\omega_1$, showing the quasi-energies of the (populated) field-dressed states. The laser frequency was $\omega_{1}=4$. The pulse shape was trapezoidal (4,1200,4) in the vector potential of amplitude $\Ahat$. For each $\alphahat$ the maximum in $R$ was renormalized to unity.     \label{fig:states1}}
\end{figure}
\begin{figure}
\includegraphics[width=0.45\textwidth]{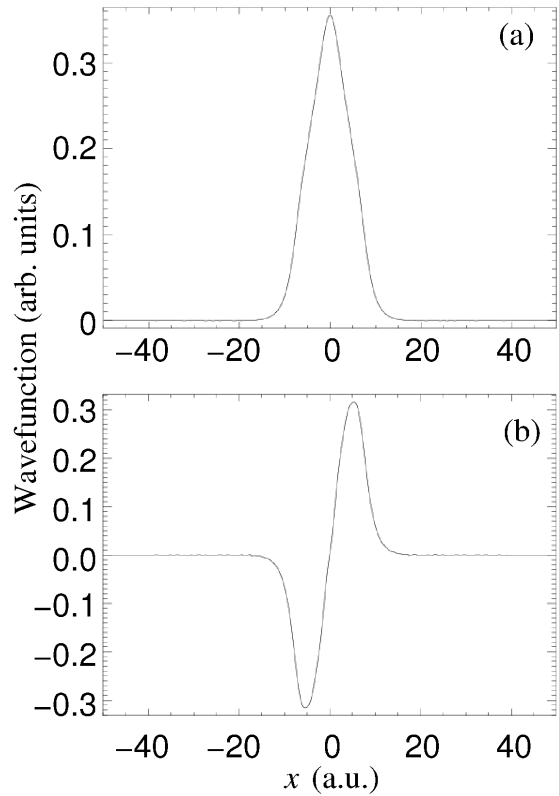} 
\caption{Field-dressed ground state wavefunction $\tilde{\varphi}_{0n}$ for  $\hat{\alpha}=4$. (a) Floquet block $n=0$, (b) $n=-1$. \label{fig:states2}}
\end{figure}

\section{Floquet state analysis of real-time wavefunctions} \label{method}
The extraction of Floquet information contained in the real-time wavefunction is useful to analyze any feature of interest in HG spectra. We start with the determination of the {(real part of the)} quasi-energy of the populated Floquet states. Once these energies are known the corresponding  Floquet states can be obtained. The method is similar to the one proposed in \cite{feitflecksteiger} for field-free dynamics.

The numerical solution of the time-dependent Schr\"odinger equation in real time yields $\Psigrid(x,t)$. Upon multiplication of \reff{psixtsuperpos} by an even or odd test function $q_\pm(x)$, spatial integration, and Fourier transformation from the time to the energy domain,{
\beq Q_\pm(\energy)=  \sum_{\beta n} \int_{t_1}^{t_2} \eulere^{-\imagi t(\epsilon_\beta +n\omega_1-\energy)}\diff t \int_{-\frac{L}{2}}^{\frac{L}{2}} q_\pm(x) \varphi_{\beta n}(x)\,\diff x,\label{Q}\eeq
$0\leq t_1 < t_2 \leq \tsim$ one can extract from the peak positions in  $\vert Q_\pm(\energy)\vert^2$ the real part of the Floquet energies 
\beq \Re\energy_{\beta n}=\Re\epsilon_\beta+n\omega_1 \eeq
belonging to even or odd Floquet states $\varphi_{\beta n}$, respectively.} The even test function is, e.g., simply unity for all {$-\frac{L}{2} < x < \frac{L}{2}$}, the odd test function may be chosen $1$ for $x>0$ and $-1$ for $x<0$. The purpose of these test functions is to extract the even and odd parity Floquet states separately. 
{Of course, only the energies of the populated (and thus relevant) Floquet states $\varphi_{\beta n}$ are obtained in this way.}

{The imaginary part $\Gamma_\beta/2$ of $\epsilon_\beta$ contributes to the width of the peaks in $\vert Q_\pm(\energy)\vert^2$. However, in our finite-time, finite-grid TDSE simulations the width of the peaks  in $\vert Q_\pm(\energy)\vert^2$ also depend on the integration time $t_2-t_1$ and the grid-size because of the absorbing grid boundaries. Only for a flat-top laser pulse and a  very long simulation time a stationary absorption rate at the grid boundaries would be established, and $\Gamma_\beta$ could be determined from the peak-width. This, however, is exactly the regime where the standard Floquet approach should be applied. We focus here on aspects of our method complementary to the conventional Floquet method, in particular  its applicability to finite pulses and time-resolved studies. }

If we multiply the wavefunction \reff{psixtsuperpos} by $\exp(\imagi t \energy)$ {(with $\energy$ real)} and integrate over time, mainly the Floquet state $\varphi_{\energy}$ for which the phase is stationary, i.e., {$\energy=\Re\epsilon_\beta+n\omega_1$} ``survives,''
\beq \varphi_{\energy}(x) \sim \int_{t_1}^{t_2} \eulere^{\imagi t \energy} \Psi(x,t) \,\diff t . \label{extract}\eeq
 The integration time $t_2-t_1$  has to be sufficiently long in order to cover  many temporal oscillations of the wavefunction.

Starting from the ground state in the potential \reff{ptpot}, we solved the TDSE for a high-frequency laser field of vector potential $A(t)= -\Ahat(t) \sin \omega_1 t$ for $\omega_1=4$ and $\Ahat(t)$ a trapezoidal pulse shape with linear up- and down-ramps over 4 cycles and 1200 cycles constant amplitude $\Ahat$ (denoted in the form (4,1200,4) in the following). Figure~\ref{fig:states1} shows 
\beq R = \vert Q_+ \vert^2 + \vert Q_- \vert^2 \label{eqR}  \eeq
(with the time-integral in \reff{Q} performed over the entire pulse) as a contour plot vs the excursion amplitude $\alphahat=\Ahat/\omega_1$ and energy $\energy$ for an energy interval within the zeroth Floquet block $n=0$. Plotting $\vert Q_+\vert^2$ and $\vert Q_-\vert^2$ individually allows to distinguish the parity of the states (labeled 'even' or 'odd' in Fig.~\ref{fig:states1}). For $\alphahat\to 0$ only the field-free state at $\energy=-0.5$ remains. However, with increasing excursion amplitude $\alphahat$ light-induced quasi-bound states emerge, which are populated due to the finite rise-time of the laser field. From the populations (see color-coding) one infers that around $\alphahat=6$ besides the field-dressed ground state the second excited field-dressed state is more populated than the first excited. For increasing $\alphahat$ the field-dressed ground state and the field-dressed first excited state become almost degenerate so that $\Delta\epsilon \to 0$ in \reff{peakpos}, {explaining the peaks at even harmonics of the laser frequency due to hyper-Raman scattering.}

Using \reff{extract} we extracted field-dressed states.
Figure~\ref{fig:states2} shows the field-dressed ground state for the Floquet blocks $n=0$ (a) and $n=-1$ (b) for $\alphahat=4$. The integration time was again the pulse duration. Equation~\reff{extract} in general yields a complex wavefunction $\varphi_\energy=\tilde{\varphi}_\energy\eulere^{\imagi \theta} $. The plots in Fig.~\ref{fig:states2} show the real wavefunction $\tilde{\varphi}_\energy$. It is seen that the parity indeed changes as one decreases $n$ by one. For $n=0$ and $\alphahat=0$ the ground state must be even. Hence, for $n=-1$ it is odd, in accordance with \reff{parity}.

It is known that if the laser frequency is tuned around resonances field-dressed states originating from different Floquet blocks (and corresponding to the coupled field-free states) display avoided crossings. These crossings have been shown to be related to localization, and to chaos in the corresponding classical system \cite{timberlakereichl}.  The separation of the two dressed states involved corresponds to the Rabi frequency and is proportional to the field strength of the driving laser. We will now show that the same is observed for transitions between already dressed states, i.e., we use the laser of frequency $\omega_1$ to dress the system and a second, weaker laser of frequency $\tilde{\omega}$ to induce transitions between dressed states. The second laser will dress the already dressed system \cite{shamailovaetal}, and the ``dressed$^2$'' states (or two color-dressed states) should display avoided crossings as the frequency $\tilde{\omega}$ is tuned around the energy gap of two dressed states.  

\begin{figure}
\includegraphics[width=0.475\textwidth]{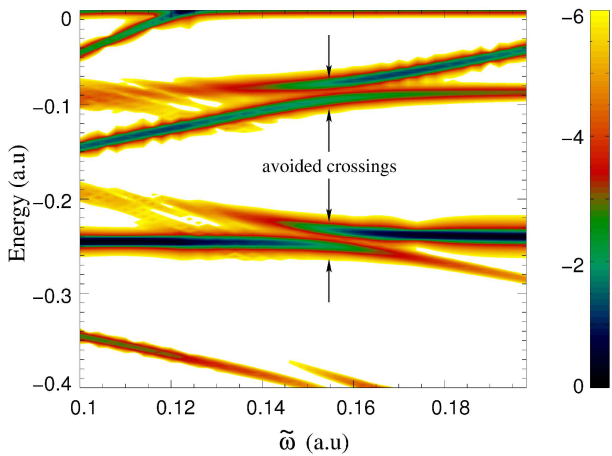} 
\caption{ (color online). $R$ vs energy $\energy$ and second-laser frequency $\tilde{\omega}$ for first-laser excursion $\hat{\alpha}=2.5$ and second-laser field strength amplitude  $\tilde{\Ehat}=0.01$. \label{fig:avcross}}
\end{figure}

From  Fig.~\ref{fig:states1} one infers that for an excursion amplitude, $\alphahat=2.5$ the energy difference between the field-dressed ground state and the field-dressed first excited state is {$\Re\epsilon_1-\Re\epsilon_0\simeq 0.155$}. Hence, we tune the frequency  $\tilde{\omega}$ of the second laser around this energy difference. The pulse envelope was the same for both lasers, and the electric field amplitude of the second laser was $\tilde{\Ehat}=0.01=\tilde{\Ahat}\tilde{\omega}=\tilde{\alphahat}\tilde{\omega}^2$ for all $\tilde{\omega}$. Figure~\ref{fig:avcross} shows results for the Floquet energy spectrum $R$ vs energy and $\tilde{\omega}$ for $\alphahat=2.5$. If the two frequencies $\omega_1$ and $\tilde{\omega}$ are incommensurate the Hamiltonian is not periodic at all.  However, our approach does not require periodicity, and we expect a Floquet analysis to be meaningful as long as the two-color Hamiltonian is {\em approximately} periodic, namely in $\tilde{T}=2\pi/\tilde{\omega}$ because $\omega_1\gg\tilde{\omega}$. In fact, the avoided crossings of {$\Re\epsilon_0$ with $\Re\epsilon_1-\tilde{\omega}$ and of  $\Re\epsilon_0+\tilde{\omega}$ with  $\Re\epsilon_1$} around $\tilde{\omega}=0.155$ are clearly visible in  Fig.~\ref{fig:avcross}.

\section{Transformations} \label{trafos}
We consider transformations $\hat{G}(t)$ which are periodic in time and reduce to unity as the laser field goes to zero, 
\beq \hat{G}(t+T)=\hat{G}(t), \qquad \hat{G}(t)\vert_{\alpha,E,A=0}=\hat{1}. \label{periodicG} \eeq
Now, since each Floquet state $\Phi_\beta$ fulfills \reff{fhami},
\beq \hat{G}(t) \hamop(t)\hat{G}^{-1}(t)\hat{G}(t) \ket{\Phi_{\beta}(t)} = \hamop'(t) \ket{\Phi'_{\beta}(t)}= \epsilon_\beta \ket{\Phi'_{\beta}(t)} \eeq
where $\hamop'(t)=\hat{G}(t)\hamop(t)\hat{G}^{-1}(t)$ is the transformed Floquet-Hamiltonian and $ \ket{\Phi'_{\beta}(t)}=\hat{G}(t) \ket{\Phi_{\beta}(t)} $
the transformed Floquet state. The quasi-energy $\epsilon_\beta$ is not affected by the transformation, and
$\ket{\Phi'_{\beta}(t)}$ is also periodic because of \reff{periodicG}, so that with \reff{fourexp}
\beq \sum_n \eulere^{-\imagi n\omega_1 t} \ket{\varphi'_{\beta n}} = \sum_{n m} \eulere^{-\imagi (n+m)\omega_1 t} \hat{G}_m \ket{\varphi_{\beta n}}, \label{vierzig}\eeq 
where $\hat{G}(t)=\sum_m \eulere^{-\imagi m \omega_1 t} \hat{G}_m$, and thus
\beq \ket{\varphi'_{\beta \ell}} = \sum_{n }  \hat{G}_{\ell-n} \ket{\varphi_{\beta n}}. \eeq 
We now specialize on transformations $\hat{G}$ that commute with the dynamical symmetry operation $\Pop_{pt}$, 
\beq [\hat{G}(t),\Pop_{pt}]=0. \label{commtrafos}\eeq
Examples are gauge transformations, e.g., for the transformation from velocity gauge, where
\beq \Wop(t)=\pop A(t) + \frac{1}{2} A^2(t), \label{vgW} \eeq
to the length gauge one has
\beq G_{\mathrm{LG}} (t) = \exp\left[ \imagi x A(t) \right]. \eeq
 Another example is the Pauli-Fierz or Kramers-Henneberger (KH) transformation, which is {\em not} a gauge transformation (although one frequently finds the term ``KH gauge'' in the literature).  If we start from the velocity gauge interaction \reff{vgW} the KH transformation reads
\beq \GKH (t) = \exp\left[\frac{\imagi}{2} \int_{\infty}^t A^2(t')\,\diff t'   + \imagi\alpha(t)\pop\right]. \eeq
This amounts to a translation in position space by the free electron excursion $\alpha(t)$  \reff{alphaintegral} and a purely time-dependent contact transformation. The KH Floquet-Hamiltonian is
\beq \hamop'(t)=\hamopKH(t) =  \halb\pop^2 + V[x+\alpha(t)] -\imagi\pabl{}{t}. \eeq
As a consequence of \reff{commtrafos}, 
\beq \Pop_{pt} \ket{\Phi_\beta'(t)} = \hat{G}(t)  \Pop_{pt} \ket{\Phi_\beta(t)} = \pm \ket{\Phi_\beta'(t)} \eeq
with the eigenvalue $\pm 1$ the same as for $ \Pop_{pt} \ket{\Phi_\beta(t)} = \pm \ket{\Phi_\beta(t)} $. One also finds
$ \hat{G}_m = (-1)^m \Pop_p \hat{G}_m \Pop_p $
and
$ \Pop_p \ket{\varphi'_{\beta \ell}} = \pm (-1)^\ell \ket{\varphi'_{\beta \ell}}$, i.e., the transformed (primed) states have the same symmetry as the original states.

\begin{figure}
\includegraphics[width=0.45\textwidth]{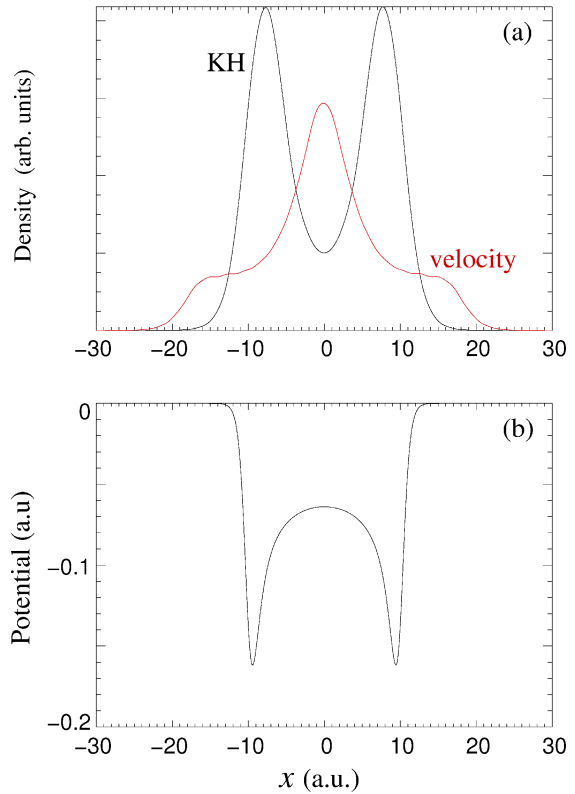} 
\caption{(a) KH and the velocity gauge probability density for the excursion amplitude $\hat{\alpha}=10$ and  target energy $\energy=-0.08$. (b) Corresponding KH potential. \label{fig:KHandvelo}}
\end{figure}

Figure~\ref{fig:KHandvelo} shows the KH and the velocity gauge probability density for the excursion amplitude $\alphahat=10$. The target energy was $\energy=-0.08$ where in Fig.~\ref{fig:states1} the almost degenerate ground and first excited state energies for $\alphahat=10$ are. The KH probability density fits to the KH potential
\beq V_\mathrm{KH}(x) = \frac{1}{2\pi} \int_0^{2\pi} V[x+\alphahat \sin \tau]\,\diff\tau, \label{KHpot}  \eeq  
shown in the lower panel. The actual calculation was performed for $\omega_1=4$ and a trapezoidal (10,1180,10)-pulse. The target energy $\energy$ in \reff{extract} is scanned through the energy region of interest, and the Floquet energy is hit when the value of the integral is maximum. If one uses the same integration time for different $\energy$ the integral 
{\beq N_\energy = \int_{-\frac{L}{2}}^\frac{L}{2} \vert\varphi_\energy(x)\vert^2 \,\diff x  \eeq } is a relative measure for the population of the {respective} Floquet state in the actual pulse.

\begin{figure}
\includegraphics[width=0.45\textwidth]{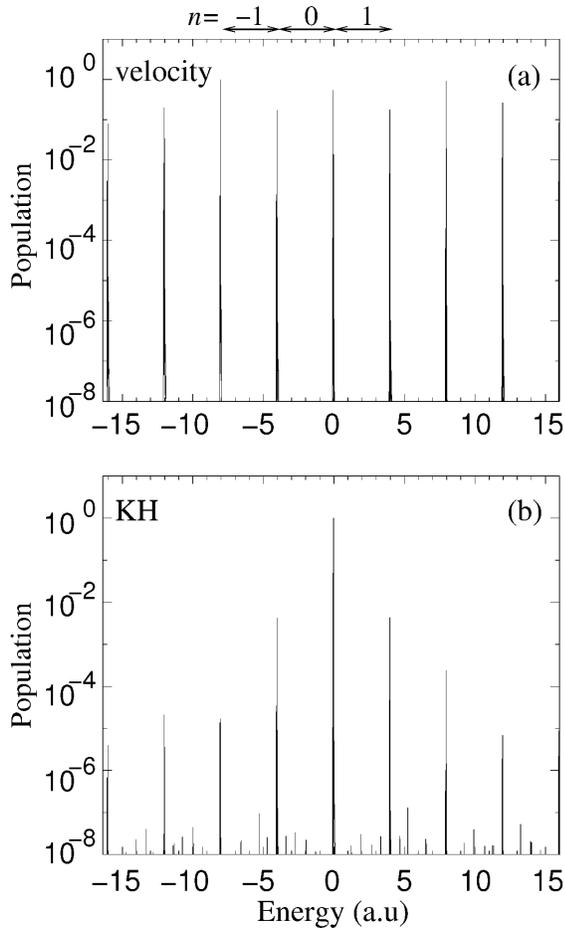} 
\caption{Floquet spectra for $\alphahat=10$, $\omega_1=4$, and a (10,1180,10)-pulse in (a) velocity gauge (with the $A^2(t)/2$-term transformed away) and (b) in the KH frame. In the KH frame the $n=0$-Floquet block dominates while in velocity gauge the population is broadly distributed over many Floquet blocks. \label{fig:popi}}
\end{figure}

The Floquet energies are invariant under the transformations $\Gop(t)$ while both the Floquet states $\ket{\varphi_{\beta n}}$ and their populations are not. In particular, in the high-frequency limit one expects that only the eigenstates in the KH potential \reff{KHpot} matter \cite{gavrila}. These states correspond to the Floquet energies in the Floquet block $n=0$. Hence, the energy spectrum in the KH frame is expected to be much more localized around $n=0$ than in velocity gauge. This is confirmed by Fig.~\ref{fig:popi}. Instead of using the even or odd test functions in \reff{Q} and spatial integration we analyzed the wavefunction $\Psigrid(x,t)$ at $x_\mathrm{test}=2$, i.e., we calculated
\beq Q'(\energy)=  \sum_{\beta n} \int_{t_1}^{t_2} \eulere^{-\imagi t(\epsilon_\beta +n\omega_1-\energy)}\diff t\  \varphi_{\beta n}(x_\mathrm{test}) \label{Qprime}.\eeq  
This avoids the transformation of the entire wavefunction to the KH frame and yields similar results as long as one chooses $x_\mathrm{test}$ in a region where the wavefunction is sizable and both odd and even parity wavefunctions contribute (for $x_\mathrm{test}=0$ only contributions from even Floquet states would be visible).
Figure~\ref{fig:popi} confirms that for transformations of the type \reff{periodicG} the populations of Floquet states  in different frames (or gauges)  are different while the Floquet energies are the same. The latter, dressed levels could be probed with a second laser \cite{morales}.
Of course, any gauge- or frame-dependence should vanish when field-free observables, such as  photoelectron spectra are considered.

\section{Photoelectron spectra} \label{photoelspec}
{Without laser field the continuum states of the P\"oschl-Teller potential have energies $\energy>0$. With laser field all continuum and bound states are contained in each Floquet block so that overlaps of dressed bound states from one block with continua from other blocks with lower $n$ are possible.} However, we expect the dressed bound states of the $n=0$ block to dominate since they are the main ones being populated during the switching-on of the laser. Let us first discuss the case where {$\omega_1>\min{\Re\epsilon_\beta}$}, i.e., a single photon is sufficient for ionization.  Then the dressed bound state in Floquet block $n$ with energy  {$\Re\epsilon_{\beta}+n\omega_1$ overlaps with continuum states of all the Floquet blocks $m<n$. In particular,  $\Re\epsilon_{\beta}+n\omega_1$} overlaps with the continuum state of energy $\epsilon_p$ of the zeroth Floquet block, where $p$ indicates the asymptotic momentum of this continuum state.

\begin{figure}
\includegraphics[width=0.45\textwidth]{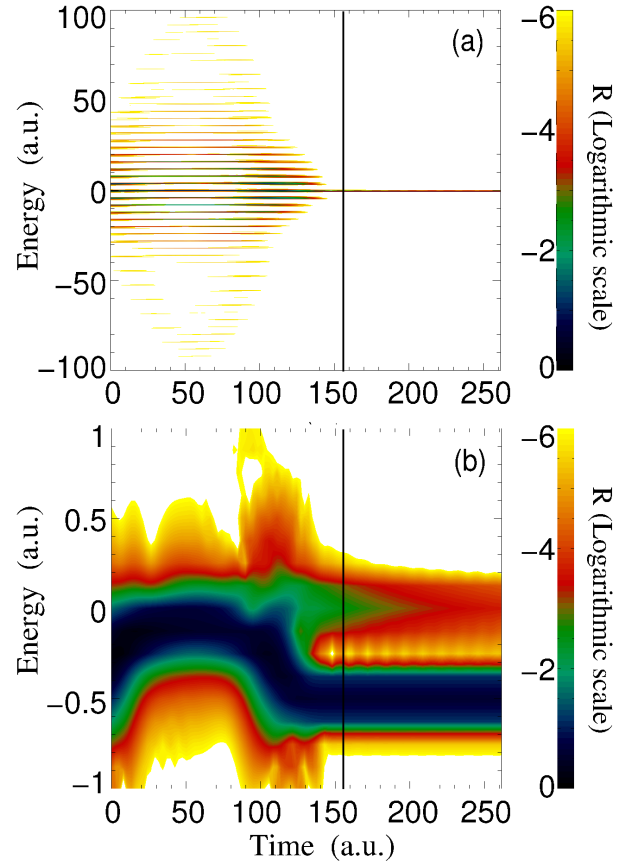} 
\caption{ (color online). Time-resolved Floquet spectra for a 100-cycle $\sin^2$-pulse of amplitude $\alphahat=\Ahat/\omega_1=10$, $\omega_1=4$, $x_\mathrm{test}=2$ (i.e., ``inside'' the potential), and a time-window of width $t_\mathrm{w}=t_2-t_1=50$. The vertical line indicates the end of the pulse. Panel (b) is a close-up of the energy region around $\epsilon^{(0)}_{0}=-0.5$ in (a). The calculation was performed in velocity gauge (with the $A^2(t)/2$-term transformed away).
\label{fig:timeresolved}}
\end{figure}

\begin{figure}
\includegraphics[width=0.45\textwidth]{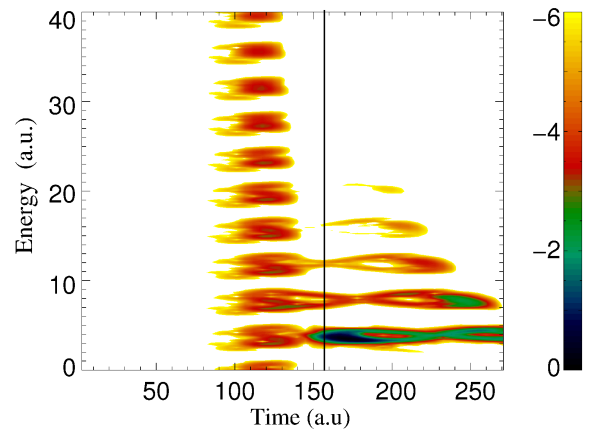} 
\caption{ (color online). Same as in Fig.~\ref{fig:timeresolved} but for $x_\mathrm{test}=471.3$.
\label{fig:timeresolved2}}
\end{figure}

We will now turn to the question of how the manifold of mixtures of bound and continuum Floquet states converts to an observable photoelectron spectrum when the pulse is switched off.  Figure~\ref{fig:timeresolved} shows a time-resolved Floquet spectrum in velocity gauge for a $\Ncyc=100$-cycle $\sin^2$-pulse
\beq A(t) = \Ahat \sin^2 \left( \frac{\omega_1 t}{2 \Ncyc} \right) \sin\omega_1 t \eeq
for $0<t<\Ncyc 2\pi/\omega_1$ and zero otherwise.
The other pulse parameters are given in the figure caption, and  $x_\mathrm{test}=2$ (i.e., ``inside'' the potential) and a time-window of width $t_\mathrm{w}=t_2-t_1=50$ were chosen for \reff{Qprime}. The time on the horizontal axis is $t_1$ so that the spectrum for times $t_1 > 100 T= 157.1$ (indicated by the vertical black line) shows field-free states, i.e.,
\begin{eqnarray} Q^{(0)}(\energy,t_1) &=&  \int_{t_1}^{t_1+t_\mathrm{w}} \eulere^{\imagi\energy t} \Psigrid(x_\mathrm{test},t)\,\diff t \label{Qnull} \\
& =& \sum_{\beta}\varphi^{(0)}_{\beta}(x_\mathrm{test}) \int_{t_1}^{t_1+t_\mathrm{w}} \eulere^{-\imagi t(\epsilon^{(0)}_\beta -\energy)}\diff t. \nonumber
\end{eqnarray}
Figure~\ref{fig:timeresolved}a shows that while the pulse is on the population is distributed over many Floquet blocks. As the pulse is switched off, all the Floquet populations for $n\neq 0$ disappear, and only the ground state population inside the potential with energy $\epsilon^{(0)}_{0}$ remains. This is because we analyzed the spectrum at the position $x_\mathrm{test}=2$. Contributions to the wavefunction corresponding to electrons in the continuum, traveling with an asymptotic  momentum $p$, decay at $x_\mathrm{test}=2$. Figure~\ref{fig:timeresolved}b shows a close-up of the region around $\epsilon^{(0)}_{0}$. With increasing amplitude of the laser pulse the dominant Floquet population shifts adiabatically from the field-free value $\epsilon^{(0)}_{0}=-0.5$ to the ground state energy of the KH potential $\epsilon^{(\mathrm{KH})}_{0}\simeq -0.09$ (see Fig.~\ref{fig:states1} for $\alphahat=10$) and back. Note that although the calculation was performed in velocity gauge the KH ground state energy is relevant here because the Floquet quasi-energies are frame- and gauge-independent.

Figure~\ref{fig:timeresolved2} shows the same analysis for $x_\mathrm{test}=471.3$, i.e., ``far away'' from the atom so that it takes some time until probability density arrives there, namely around $t=100$. It is interesting to observe that in velocity gauge this ``arrival time'' during the pulse is independent of the energy.  As the laser pulse is switched off at $t=157.1$ many Floquet channels close. However, because electrons are still on their way from the atom to the ``virtual detector'' at $x_\mathrm{test}=471.3$ we are able to ``measure'' the field-free photoelectron spectrum of the electrons emitted in that direction. The time these free electrons need to pass the virtual detector decreases with increasing energy, as is seen in Fig.~\ref{fig:timeresolved2} where the width of the traces for $t>157.1$ decrease   with increasing energy. The five traces visible are separated by $\omega_1$ and correspond to above-threshold ionization (ATI) peaks (see, e.g., the review\cite{ATI} or \cite{lowfrequstuff}). They are quite broad in energy because of the change of the ionization potential (from field-free value to KH value and back). Their figure-eight shape in the contour plot of  Fig.~\ref{fig:timeresolved2} is a peculiarity of the $\sin^2$-pulse shape.

\begin{figure}
\includegraphics[width=0.45\textwidth]{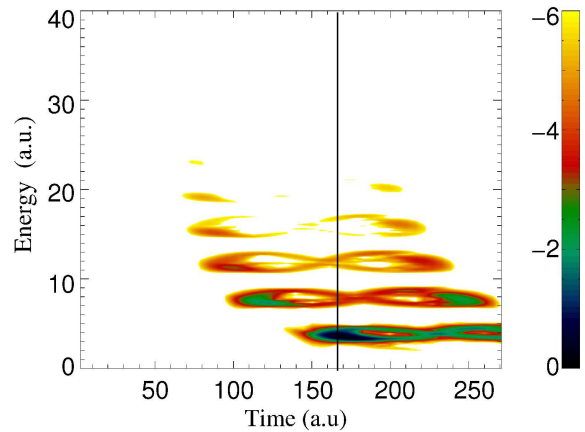} 
\caption{ (color online). Same as Fig.~\ref{fig:timeresolved2} but in the KH frame. \label{fig:kh}}
\end{figure}

Figure~\ref{fig:kh} shows the corresponding result obtained in the KH frame. We see that in the KH frame only those states are populated in the laser field which actually contribute to the final field-free spectrum. This is because the KH potential at  $x_\mathrm{test}=471.3$ is almost identical to the field-free one so that outgoing electrons are not affected anymore by the oscillating KH binding potential. It is also seen in Fig.~\ref{fig:kh} that the most energetic electrons arrive earlier at $x_\mathrm{test}$, unlike the velocity gauge-result in Fig.~\ref{fig:timeresolved2}.

\section{Channel-closings} \label{channelclose}
So far we studied mainly high-frequency phenomena where the Floquet blocks are well separated on the atomic energy scale because the laser frequency exceeds the ground state ionization potential. However, there are plenty of interesting, non-perturbative phenomena occurring at low frequencies where the ponderomotive energy $\Up=\Ehat^2/4\omega_1^2$ can be large at nowadays available laser intensities $\Ehat^2$. Examples are tunneling ionization and high-order ATI due to rescattering of electrons \cite{lowfrequstuff,highorderATI}. In this Section we choose the so-called ``channel-closing'' (see \cite{channelclosings} and references therein) as a low-frequency phenomenon to illustrate our method.  

\begin{figure}
\includegraphics[width=0.45\textwidth]{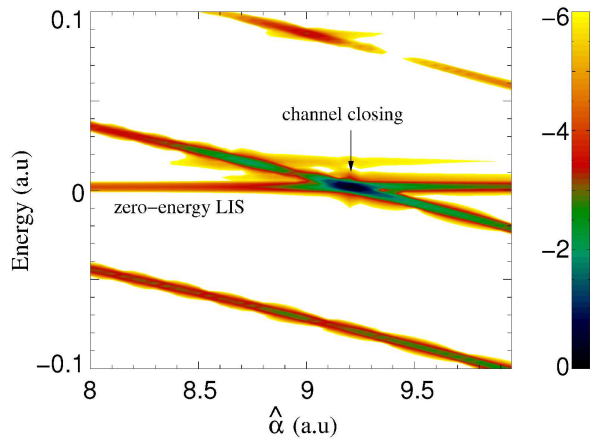} 
\caption{ (color online). Logarithmic plot of $ R = \vert Q_+\vert^2 + \vert Q_- \vert^2 $ vs energy $\energy$ and excursion amplitude $\alphahat$, showing the (populated) field-dressed states for $\omega_{1}=0.08$ and a trapezoidal (4,40,4)-pulse.  \label{fig:lowfrequ}}
\end{figure}

The TDSE was solved for a trapezoidal pulse of frequency $\omega_1=0.08$.  On the energy scale of the ionization potential the Floquet blocks are packed much closer in this case, meaning that many photons are necessary for ionization. In Fig.~\ref{fig:lowfrequ} we plot the Floquet energy spectrum $R$ in a certain range of excursion amplitude $\alphahat=\Ehat/\omega_1^2$ and energy $\energy$ around the field-free continuum threshold (other relevant parameters given in the figure caption).  The calculation was performed in velocity gauge using again the potential \reff{ptpot}.   There is a clear down-shift of all the populated Floquet levels with increasing laser amplitude. This AC Stark shift is also referred to as the ``ponderomotive shift'' because the effective ionization potential is increased by $\Up$. In fact, the energy in the photoelectron spectrum is given by \beq \energy= \frac{p^2}{2} =n\hbar\omega_1-(\arrowvert\energy_0 \arrowvert +\Up), \label{closings}\eeq
(provided the AC Stark shift of the initial state is negligible, which for atomic ground states at long wavelengths often is the case). 
$\energy_{0}$ is the initial electron energy and $n$ is the number of photons absorbed. In order to reach the continuum at all $n > (\arrowvert\energy_0 \arrowvert +\Up)/\hbar \omega_1$ photons have to be absorbed.  As the intensity, and thus $\Up$, is increased, more and more photons are needed for ionization. When $n$ photons are no longer sufficient but $n+1$ photons are needed the $n$-photon  channel closing occurs. In the plot shown in Fig.~\ref{fig:lowfrequ} a channel closing manifests itself as a crossing of a Floquet quasi-energy and the continuum threshold. Now, the interesting feature in  Fig.~\ref{fig:lowfrequ}  is the zero-energy LIS. Such LIS were also observed in Ref.~\cite{wassaf}, where  their connection with experimentally observed enhancements in the photoelectron spectra at high energies \cite{hertleinandothers} was established. The parity of both states involved in the crossing in Fig.~\ref{fig:lowfrequ} is even, and it is known that depending on the parity of the states, channel closings affect the photoelectron spectrum differently \cite{wassaf,popru}.

\begin{figure}
\includegraphics[width=0.45\textwidth]{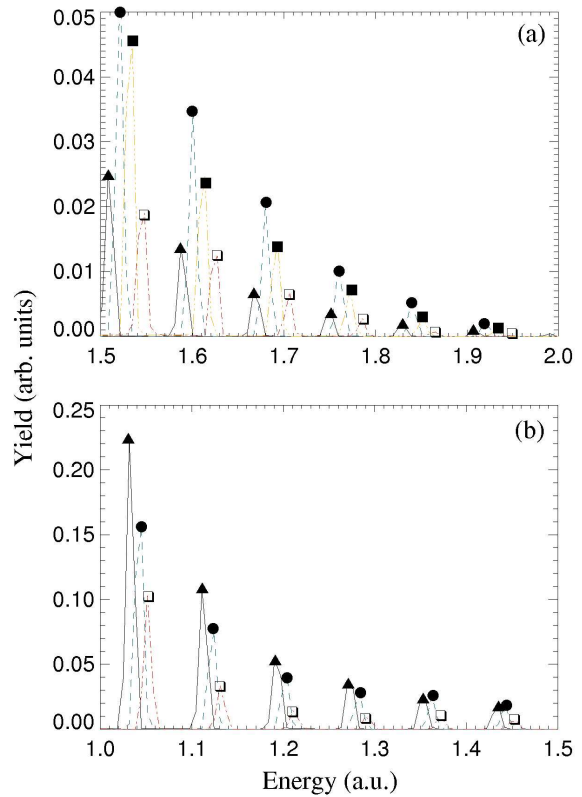} 
\caption{ (color online).  Photoelectron spectra around  $5\Up$, parameters as in Fig.~\ref{fig:lowfrequ}. (a) Non-monotonic behavior of the yield (open squares $\alphahat=13.0$, solid squares $13.3$, circles $13.55$, triangles $13.8$). (b) Same for an odd channel closing, showing a  monotonic behavior of the yield with increasing intensity (triangles $\alpha= 11.8$, circles $11.55$, open squares $\alpha= 11.3$).   \label{fig:enhancment}}
\end{figure}

  In our model, for the first even channel closing eight photons are needed. According \reff{closings} it is expected at $\alphahat=9.354$, which indeed is close to where the crossing is observed in Fig.~\ref{fig:lowfrequ}. The small discrepancy is because of the AC Stark shift of the initial state, neglected in 
\reff{closings}. One would expect that channel closings only affect low-energy electrons because the kinetic energy of the electrons whose channel is about to close is low.  Hence, as the intensity is increased the yield of ATI peaks at energies, say, $> 5\Up$ should increase monotonously as well.  However, near even photon channel closing there is a marked increase in the photoelectron yield at high energies \cite{popru,wassaf,channelclosings}. Instead,  when in odd photon channel closings the odd-parity LIS crosses the zero-energy LIS, such enhancements are absent. The first odd photon channel closing occurs around $\alphahat=11.55$, the next even photon channel closing occurs around $\alphahat=13.55$. The photoelectron spectra obtained using our Floquet method confirm the presence and absence of enhancements at even and odd channel closings, respectively, as shown in Fig~\ref{fig:enhancment}.

\section{Conclusions} \label{sumout}
We described a method for obtaining Floquet information from real-time wavefunctions. In this approach, it is not necessary to assume strict periodicity. In fact, it is possible  to follow the time-resolved Floquet quasi-energies as they shift during a laser pulse. Moreover, the populations of the Floquet states can be determined so that especially cases where superpositions of Floquet states play a role can be identified. The usefulness of the method was illustrated by several examples employing the one-dimensional P\"oschl-Teller potential with only a single field-free {bound} state. In particular, the origin of peaks at even harmonics of the laser frequency in an inversion-symmetric potential, avoided crossings of dressed already field-dressed states induced by a second laser, the properties of Floquet states under time-periodic transformations, the emergence of invariant, observable photoelectron spectra after the laser pulse, and photoelectron enhancements at channel closings were discussed. The method is straightforwardly extendable to three dimensions. We think the method is most useful for researchers running codes to solve the time-dependent Schr\"odinger equation in real time. By saving the wavefunction at selected spatial positions as a function of time during the interaction with the laser field the analysis in terms of light-induced states can be easily performed {\em a posteriori}. The application of the method to correlated multi-electron systems may be very fruitful, as the understanding of field-dressed, multiply-excited or autoionizing states is still poor.

\section*{Acknowledgment}

This work was supported by the SFB 652 of the German Science Foundation (DFG).

\begin{appendix}

\section{Extension to three dimensions} \label{3D}
The method of Floquet analysis described in this work is easily extendable to higher-dimensional systems. For, e.g.,  hydrogenic systems in 3 dimensions (3D) one could follow the evolution of Floquet states with different orbital angular momentum quantum numbers $l$. Instead of \reff{psixt} we have
 \beq
 \Psi(r,\theta,\phi,t)=e^{-\imagi\epsilon t}\Phi(r,\theta,\phi,t)
 \eeq
with [compare to \reff{fourexp}]
 \beq
 \Phi(r,\theta,\phi,t)=\sum_{n} \varphi_{n}(r,\theta,\phi) \eulere^{-\imagi n\omega_1 t} .
 \eeq
The operator $\Pop_{pt}$ [compare to \reff{poppt}] acts according
\beq \Pop_{pt} f(\vektr,t)=f(-\vektr,t+\frac{\pi}{\omega_1}),  \eeq
and \reff{alternparity} becomes
\beq
\Pop_{pt}\Phi(r,\theta,\phi,t)=\sum_{n}(-1)^{n}\exp(-\imagi n\omega_1 t)\Pop_{p}\varphi_{n}(r,\theta,\phi).
\eeq
If we expand the $\varphi_{n}(r,\theta,\phi)$ in spherical harmonics,
\beq \varphi_{n}(r,\theta,\phi) = R_{nl}(r)Y_{lm}(\theta,\phi), \eeq
we find, using
\beq \Pop_{p}Y_{lm}(\theta,\phi)= Y_{lm}(\pi-\theta,\pi+\phi) =(-1)^l Y_{lm}(\theta,\phi),\eeq
that
\beq
\Pop_{p}\varphi_{n}(r,\theta,\phi)=(-1)^{n+l}\varphi_{n}(r,\theta,\phi),
\eeq
the analogue of \reff{parity}. Note that $n$ is the Floquet block index here, not the principal quantum number. After these considerations it is straightforward to extend  the Floquet analysis of real-time wavefunctions described in this work to 3D.

\end{appendix}

\end{document}